\documentclass[final]{elsarticle}
\pdfoutput=1
\usepackage{hyperref}
\usepackage{lineno,color}
\journal{NIM-A}
\usepackage{amsmath}
\usepackage{graphicx}
\usepackage[left=2cm,right=2cm,top=2cm,bottom=2cm]{geometry}

\newcommand\aap{A\&A}
\newcommand\apjl{ApJ}

\def\procspie{\ref@jnl{Proc.~SPIE}}   % Proceedings of the SPIE

\usepackage{float}
\usepackage{eurosym}
\usepackage{subcaption}
\usepackage{multirow}

\biboptions{authoryear}
\begin{document}

\begin{frontmatter}

\title{The Light-Trap: A novel concept for a large SiPM-based pixel for Very High Energy gamma-ray astronomy and beyond}

\author[a,z]{D.~Guberman}
\ead{daniel.guberman@pi.infn.it}
\author[a,y]{J.~Cortina}
\author[a]{J.~E.~Ward}
\author[a]{E.~Do Souto Espi\~nera}
\author[b]{A.~Hahn}
\author[b,c]{D.~Mazin}

\address[a]{Institut de F\'isica d'Altes Energies (IFAE), The Barcelona Institute of Science and Technology (BIST), E-08193 Bellaterra (Barcelona), Spain}
\address[b]{Max-Planck-Institut f\"ur Physik, D-80805 M\"unchen, Germany}
\address[c]{Institute for Cosmic Ray Research, University of Tokyo Kashiwa-no-ha 5-1-5, Kashiwa-shi, 277-8582 Chiba, Japan}
\address[z]{Now at Istituto Nazionale di Fisica Nucleare (INFN) Pisa, I-56126 Pisa, Italy}
\address[y]{Now at Centro de Investigaciones Energéticas, Medioambientales y Tecnológicas (CIEMAT), Avda. Complutense 40, E-28040 Madrid, Spain}


\begin{abstract}

Among the main disadvantages of using silicon photomultipliers (SiPMs) in large experiments are their limited physical area (increasing the cost and the complexity of the readout of a camera) and their sensitivity to unwanted wavelengths. This explains why photomultiplier tubes (PMTs) are still selected for the largest cameras of present and future Very High Energy (VHE) gamma-ray telescopes. These telescopes require photosensors that are sensitive to the fast and dim optical/near-UV Cherenkov radiation emitted due to the interaction of gamma rays with the atmosphere. Here we introduce a low-cost pixel consisting of a SiPM attached to a PMMA disk doped with a wavelength-shifting material, which collects light over a much larger area than standard SiPMs, increases sensitivity to near-UV light and improves background rejection. We also show the measurements performed in the laboratory with a proof-of-concept \textit{Light-Trap} pixel that is equipped with a 3$\times$3~mm$^2$ SiPM collecting light only in the 300-400~nm band, covering an area $\sim$20 times larger than that of the same SiPM itself. We also present results from simulations performed with Geant4 to evaluate its performance. In addition to VHE astronomy, this pixel could have other applications in fields where detection area and cost are critical. 
\end{abstract}


\end{frontmatter}

%\linenumbers

\section{Introduction}
\label{sec:intro}

The Imaging Atmospheric Cherenkov Technique (IACT) uses one or several telescopes to record the Cherenkov light flashes produced by extensive air showers initiated by Very High Energy (VHE, E $>$ 50 GeV) gamma rays in the upper atmosphere. These telescopes typically feature 1-30~m diameter mirrors and cameras of a few hundreds to a few thousands of pixels. The detection of the Cherenkov flashes is challenging because of their short duration (a few nanoseconds) and their low light intensity (down to a few phe per pixel). The IACT has been relying on photomultiplier tubes (PMTs) for this task since the early days of the technique. The three most sensitive currently operating instruments, VERITAS \citep{VERITAS2008}, H.E.S.S.\citep{HESS2006} and MAGIC \citep{upgrade1}, are equipped with PMTs that output pulses of a few ns~FWHM. However, the recent developments in silicon photomultipliers (SiPMs) threatens the hegemony of PMTs in the field. FACT is the first telescope fully equipped with SiPMs and has already been operating for several years\citep{FactMoon}. The use of SiPMs is also being considered for the next generation CTA~\citep{Acharya_2013} observatory, with several studies currently ongoing~\citep{SiPM-SST1M,SiPM-SCT,SiPM-LST,SiPM-ASTRI}.

Compared to PMTs, SiPMs provide higher photo-detection efficiency (PDE), are robust devices that do not experience any significant aging when exposed to bright environments, and do not operate under high voltage. Over the last years the performance of SiPMs has significantly improved while their costs have decreased. This is of particular interest for some of the future physics goals in VHE astronomy, which will necessitate the ability of IACT telescope systems to observe several-degree sections of the sky at one time~\citep{Acharya_2013, Machete_2016}. This demands the construction of more telescopes and/or larger cameras, needing many thousands of pixels, which rapidly becomes prohibitive if utilising PMTs at $>$ \EUR{100} /pixel.

SiPMs are however still not perfectly suitable for VHE astronomy. Typical commercial devices are not available in sizes larger than 6$\times$6 mm$^{2}$, which can be a problem when trying to build large cameras, not only because of the cost, but also because of the complexity of the readout. Building larger SiPMs is normally not considered as a feasible solution, because thermal noise (in the form of dark count rates) and specially capacitance significantly increase with size. One promising approach is to build pixels made of several SiPMs ($\sim$10) tiled together, where the signal output is the sum of all the individual signals of each SiPM~\citep{Ambrosi_2016, Hahn_2017}. Some drawbacks of this approach are that the gain of the individual SiPMs must be kept very well under control (ideally all the tiled devices should have the same gain) and that the noise of all the devices is also being summed, which disturbs the single-photon resolution and can be particularly disturbing during the calibration process.

\begin{figure}[t]
\centering
\includegraphics[width=0.7\columnwidth]{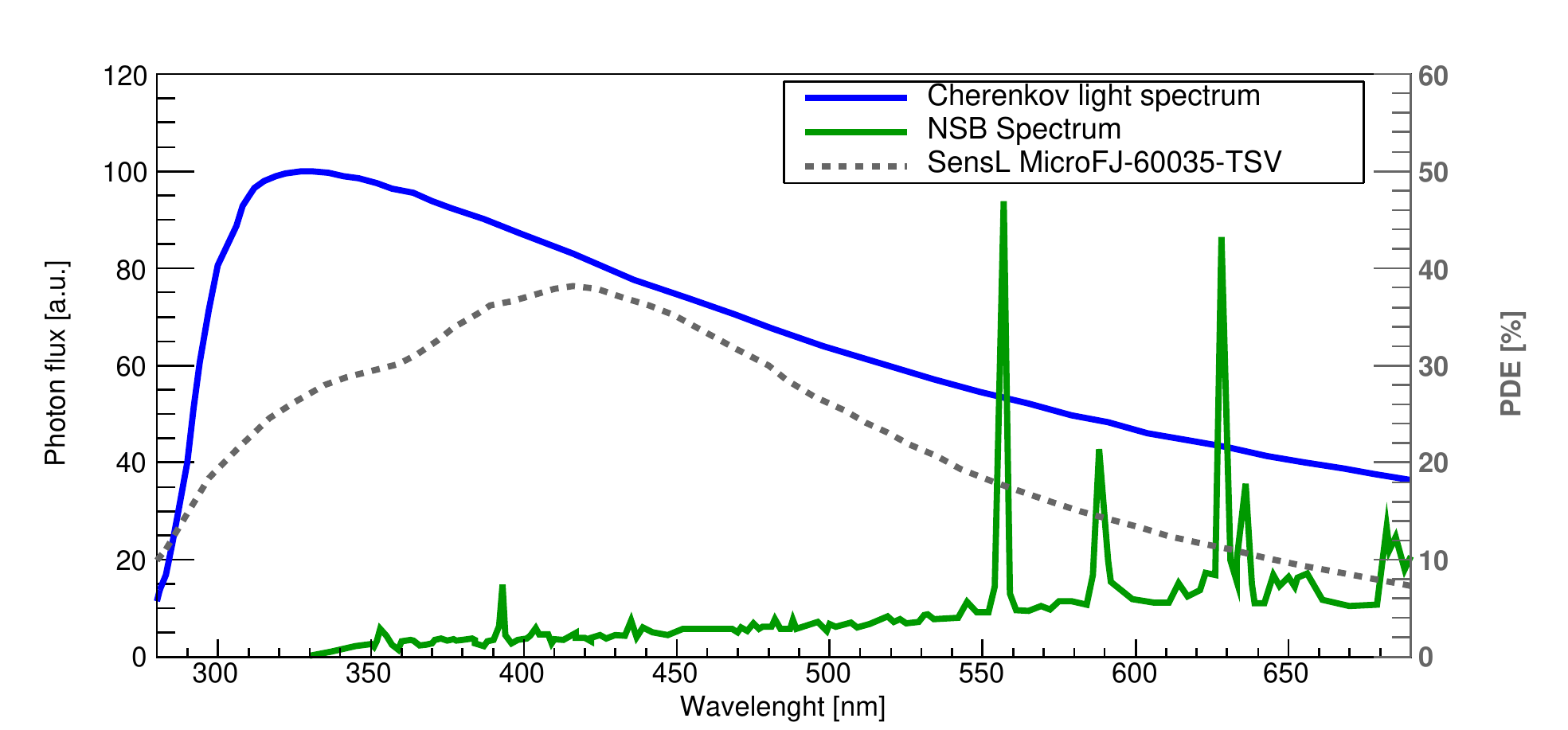}
\caption[Night Sky Background (NSB, \citealp{Benn98}) and Cherenkov light spectrum for a vertical shower initiated by a 1~TeV gamma ray~\citep{1TeVshower} at 2200 meter altitude in La Palma, Spain. As a reference, the photo-detection efficiency (PDE) of a SensL MicroFJ-60035-TSV SiPM is plotted in gray (using the right-hand axis).]{Night Sky Background (NSB, \citealt{Benn98}) and Cherenkov light spectrum for a vertical shower initiated by a 1~TeV gamma ray~\citep{1TeVshower} at 2200 meter altitude in La Palma, Spain. As a reference, the photo-detection efficiency (PDE) of a SensL MicroFJ-60035-TSV SiPM\protect\footnotemark is plotted in gray (using the right-hand axis).}\label{fig:NSB}
\end{figure}

The most common SiPMs also have the disadvantage of offering a photo-detection efficiency (PDE) that is not very sensitive below 400~nm, where most of the Cherenkov light is emitted, but too sensitive at higher wavelengths where the contribution from the Night Sky Background (NSB) is much larger (see Figure~\ref{fig:NSB}). This may however change with the recent development of SiPMs with enhanced sensitivity in the near UV band~\citep{UVFBK_2017}.

Finally, even if SiPMs are becoming cheaper, they still cost $>$1 \euro/mm$^{2}$. Hence, their cost is comparable to PMTs for devices with a detection area beyond 1~cm$^2$.

In this work we present the \textit{Light-Trap} pixel, an alternative cost-effective solution to use SiPM technology for large detection areas. In the following sections we describe the basic principles of this solution and present the results of Monte Carlo simulations and laboratory measurements performed with a prototype we designed and built as a proof-of-concept pixel.

\footnotetext{www.sensl.com}
\section{The Light-Trap pixel principles}\label{sec:LT_poc}

\begin{figure}[t]
  % Fixed length
  \centering
  \subcaptionbox{Top down view\label{fig:LTTop}}{\includegraphics[width=0.49\columnwidth]{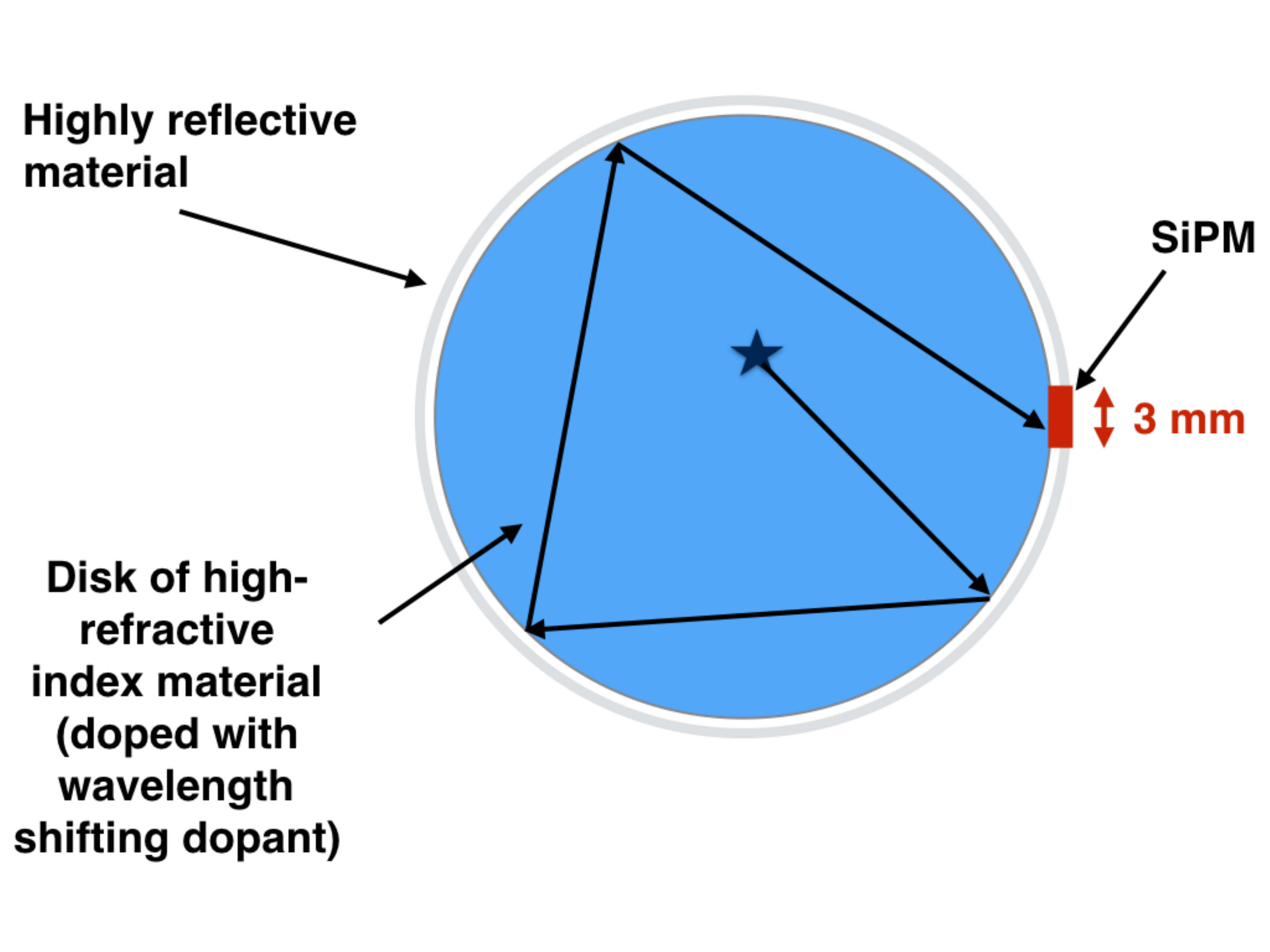}}\hspace{-0.0em}
  \subcaptionbox{Side view\label{fig:LTSide}}{\includegraphics[width=0.49\columnwidth]{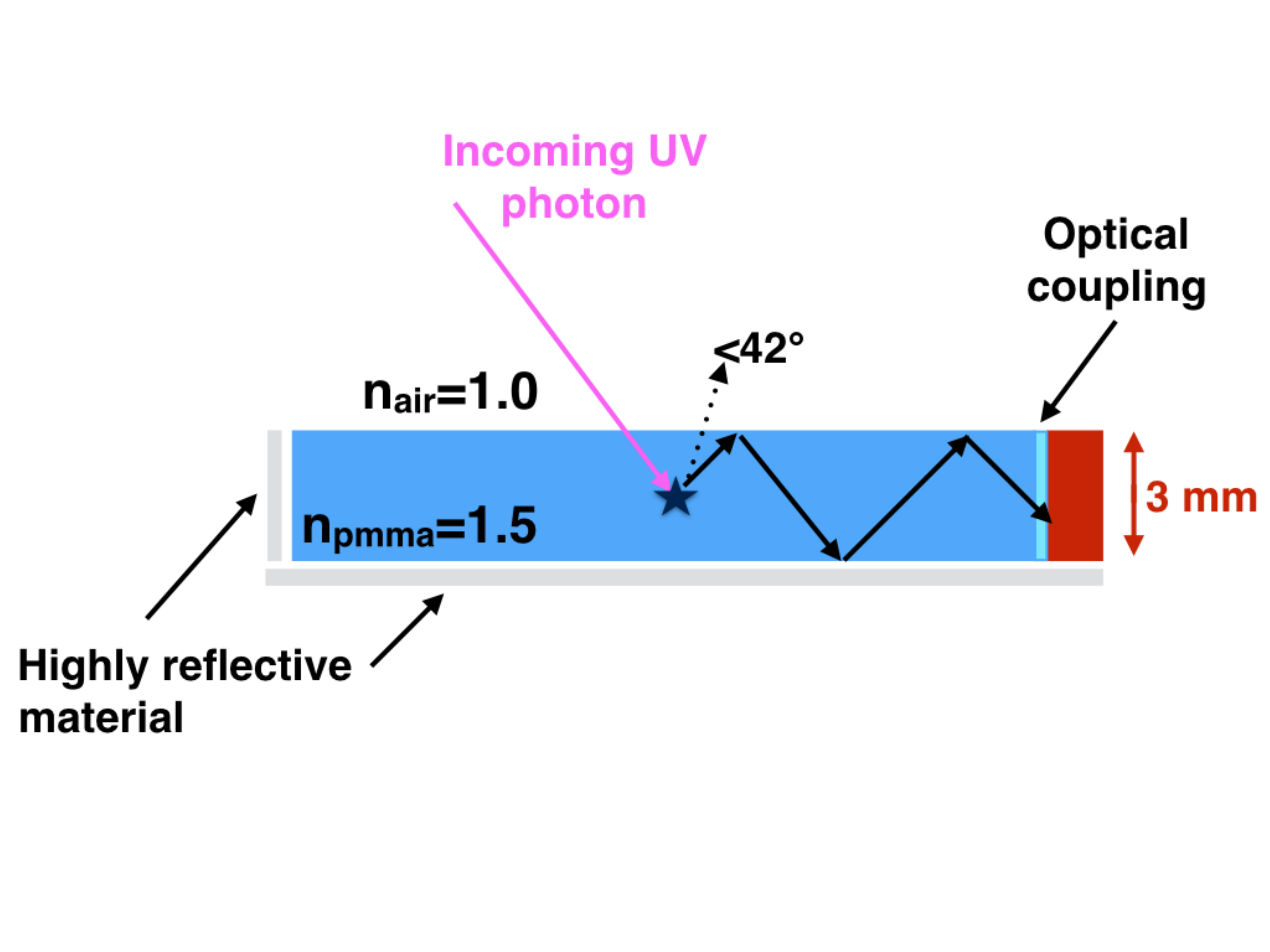}}
  \caption{Conceptual design of the Light-Trap.}
  \label{fig:LTconcept}
\end{figure}

The Light-Trap pixel consists on a SiPM coupled to a polymethylmethacrylate (PMMA) disk (see Figure \ref{fig:LTconcept}). This disk is doped with a wavelength-shifting (WLS) fluor that absorbs photons in the \mbox{$\sim$300-400 nm} wavelength range and re-emits them in the $\sim$400-500 nm range. WLS photons are re-emitted isotropically, so a fraction of them gets trapped in the disk by total internal reflection (TIR) and eventually reaches the SiPM. The rest of the re-emitted photons escape. Some of the photons that escape through the side-walls or the bottom can be recovered with the addition of reflective surfaces near the disk. There should be a minimum air gap between disk and the reflective surfaces in order not to lose the possibility of TIR at the disk surface.

As a result,

\begin{enumerate}
  \item Light around the peak of the Cherenkov spectrum ($\sim$350 nm) is collected.
  \item Light at longer wavelengths (for which the NSB dominates) is not absorbed by the WLS material and rarely reaches the SiPM.
  \item The absorbed Cherenkov photons are re-emitted at a wavelength where the SiPM PDE is higher.
  \item The collection area of the detector can be a factor $\sim$10-50 larger than the sensitive area of the SiPM, i.e. the cost is reduced by the same factor (if the cost of the disk is low) and thus enables us to build pixels far larger than commercially available SiPMs.
\end{enumerate}

\section{Proof-of-Concept pixel}\label{sec:pixel_poc}

\begin{figure}
  \centering
  \includegraphics[width=0.4\columnwidth]{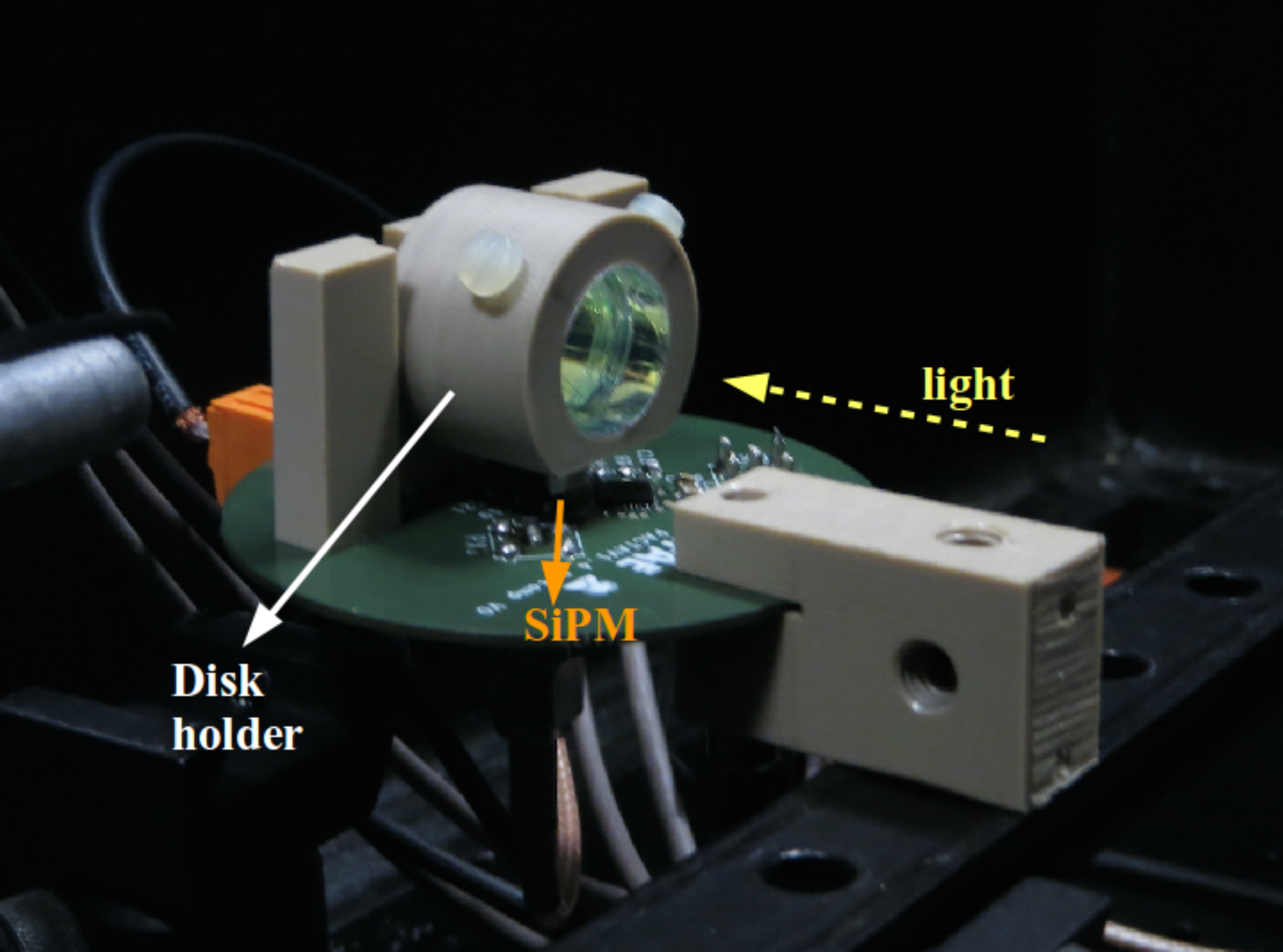} % requires the graphicx package
   \caption{Image of the PoC pixel, showing the Light-Trap mechanical holder and PCB. The SiPM sits on the PCB and looks upwards into the holder.}
   \label{fig:LTHold}
\end{figure}

Following the principles stated before, we produced a proof-of-concept (PoC) detector suitable for laboratory testing (see Figure~\ref{fig:LTHold}). A 3$\times$3~mm$^2$ SiPM is optically coupled to a 15~mm diameter and 3~mm thick PMMA disk. The detector is contained within a cylindrical polyethylene holder. The inner surface of the holder is covered with a reflective foil. In this section we discuss the individual components of the device.

\subsection{SiPM}\label{sec:ketek}

The SiPM used for this study was the 3~mm~$\times$~3~mm KETEK PM3375 (with a peak sensitivity at 420 nm\footnote{www.ketek.net}. Even if it is not the optimal device for IACT applications, mainly due to its high cross-talk probability ($\sim$36\% at 15\% over-voltage), it was suitable for our testing purposes, essentially thanks to its robust pins, which made easy the task of mounting and unmounting the SiPM from the printed circuit board (PCB) used for readout. We were not interested in the absolute properties of the sensor itself but on how the performance achieved with the Light-Trap compares to the performance of the same, \textit{naked}, SiPM used to build it.

The device we used has a breakdown voltage of $\sim$22~V at 22$^\circ$C. The tests were performed at $\sim$9~\% over-voltage. By recording events triggered only by dark counts, we estimated the cross-talk probability to be of $\sim$20\%. No temperature-control system was used to keep stable the SiPM temperature. Instead, we monitored the ambient temperature and the SiPM gain by constantly calibrating the amplitude of the single-photoelectron pulse using dark count events.

\subsection{Disk}\label{sec:disk}

\begin{figure}
\centering
\includegraphics[clip=true, trim= 0cm 0cm 0cm 0cm, angle=90, ,height=5cm]{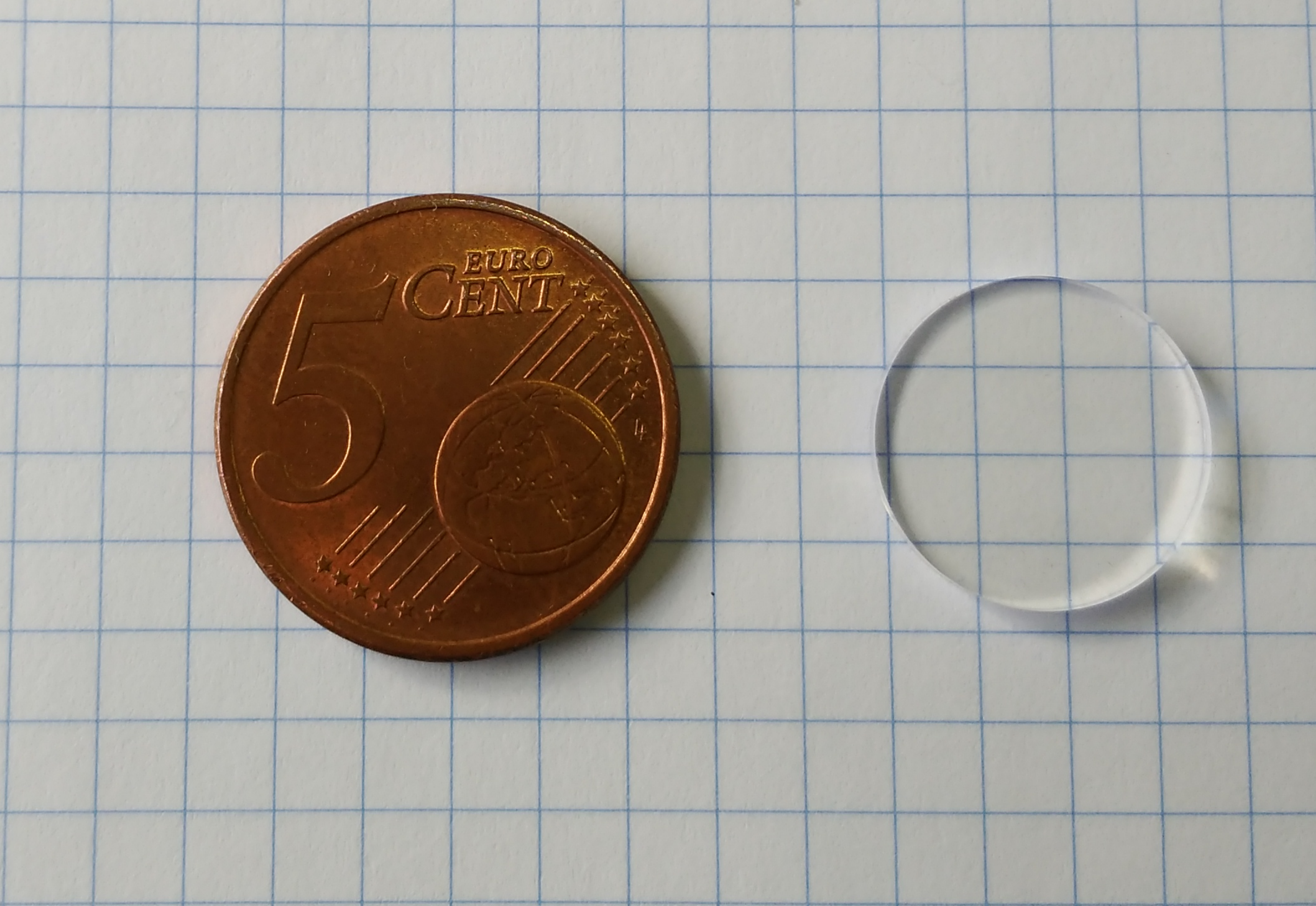}
\includegraphics[height=5cm]{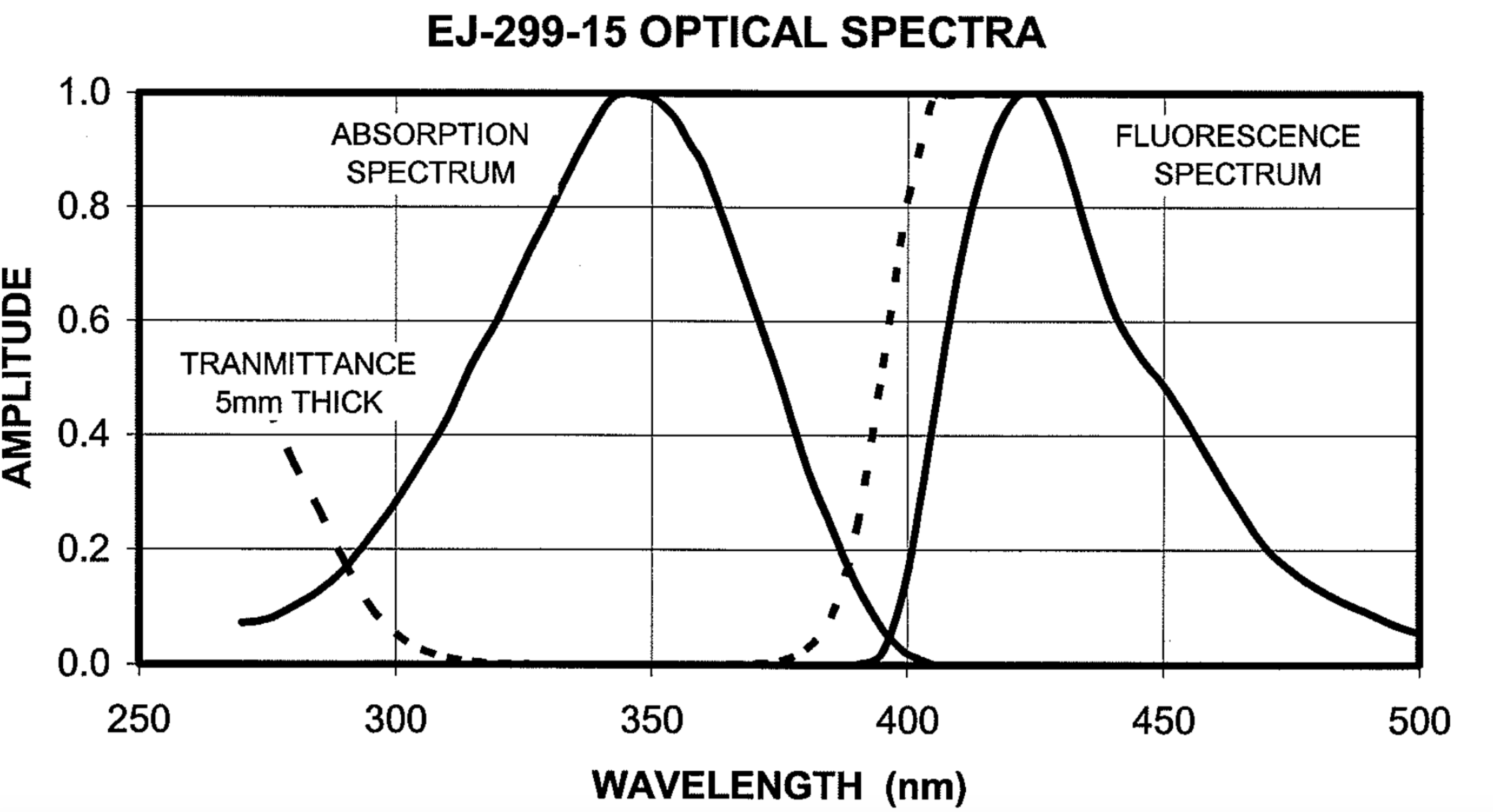}
  \caption{\textbf{Left:} PMMA disk doped with EJ-299-15 manufactured by Eljen. \textbf{Right:} EJ-299-15 absorption and emission spectra, as provided by the company.}\label{fig:EJ29915}
\end{figure}

Wavelength-shifting custom-doped polymethylmethacrylate (PMMA, refractive index $n=1.49$) disks were purchased from Eljen Technology (Sweetwater, Texas). These disks had a diameter of 15~mm and a thickness of 3~mm and were doped with EJ-299-15. They are intended to absorb light in the UV band and re-emit it by fluorescence in blue wavelengths (see Figure \ref{fig:EJ29915}). The dopant levels were customised by Eljen according to our specifications, i.e. to absorb practically 100\% of incident 340~nm photons within 1.5~mm of the material. The wavelength-shifting fluor has fast re-emission time on the order of $\sim$1 ns, with a quantum yield of $\sim$84\%. 

The manufacturing process of the disks begins with the casting of small rods of the doped plastic. The rods are then heat pressed to the desired thickness (3~mm) and the disk surfaces are manually polished using a diamond tool. This process can sometimes leave some ``footprints'' on the disks that could scatter the light reaching the disk surface, affecting the TIR efficiency. We tested the quality of the optical polishing using a green ($\sim$532~nm) and a red ($\sim$650~nm) laser, wavelengths that are not absorbed by the WLS, and measuring the light transmitted when the beams go through the disk. No significant losses due to roughness in the surface were found.

\subsection{Optical coupling}\label{sec:coupling}

Ideally, we would like that all the photons re-emitted by the WLS were reflected by the disk walls, get trapped and only be able to escape when they are approaching the SiPM. In an optimal Light-Trap pixel the outer layer of the SiPM would have the same refractive index of the disk, so that no photons approaching the detector are reflected, either by the disk wall or by the SiPM (as if the detector would be embedded in the disk). How close to this ideal situation is our real detector depends on how efficiently the disk and the SiPM are optically coupled. Our Light-Trap pixel uses an optically clear silicone rubber sheet (EJ-560, $n=1.43$, also purchased from Eljen Technology) that was cut to match the SiPM size. This silicone material is soft and only lightly adhesive, thus allowing the removal and addition of the SiPM. It features $\sim$100\% internal transmission. When no pressure is applied it has a thickness of 1~mm, that shrinks to $\sim$0.8~mm when pressed against the disk (see Section~\ref{sec:holder}).

\subsection{Reflective walls}\label{sec:mirrors}

To help improve efficiency in the case that wavelength-shifted photons do not undergo TIR, 3M\textsuperscript{\textregistered} ERS reflective foil with a thickness of $(0.14 \pm 0.01)$~mm was cut so as to surround the back and sides of the disk. The reflectivity of this material to $\sim$532~nm light was estimated to be of $(98 \pm 2)\%$, with no major dependence on the incident angle~\citep{Adrianna}. Those measurements are compatible with the reflectivity reported in Figure 5 of~\citep{Okomura_2017}.

\subsection{Pixel holder and readout electronics}\label{sec:holder}

In order to hold the SiPM, PMMA disk, and reflective foil together it was necessary to construct a cylindrical polyethylene holder. The holder inner diameter is slightly larger than the disk diameter, $(15.45 \pm 0.02)$~mm, in order to fit the disk, the reflective foil and leave some air gap in between. It should be noted that the reflective foil was firmly positioned inside this holder, although it is still expected to be in contact with the disk in some places (thus impacting TIR efficiency and the geometry of the reflections). Two plastic screws are used to apply pressure between the disk and the SiPM to improve the efficiency of the optical coupling. The holder is screwed to the PCB containing the SiPM readout electronics. The output signal of the SiPM is pre-amplified using a wideband current mode preamplifier named \textit{PACTA}~\citep{pacta_2012}, initially designed to be used with PMTs in CTA.

\section{Laboratory measurements}\label{sec:lab}

\begin{figure}
    \centering
    \includegraphics[clip=true,trim=0cm 8cm 0cm 8cm,width=0.48\textwidth]{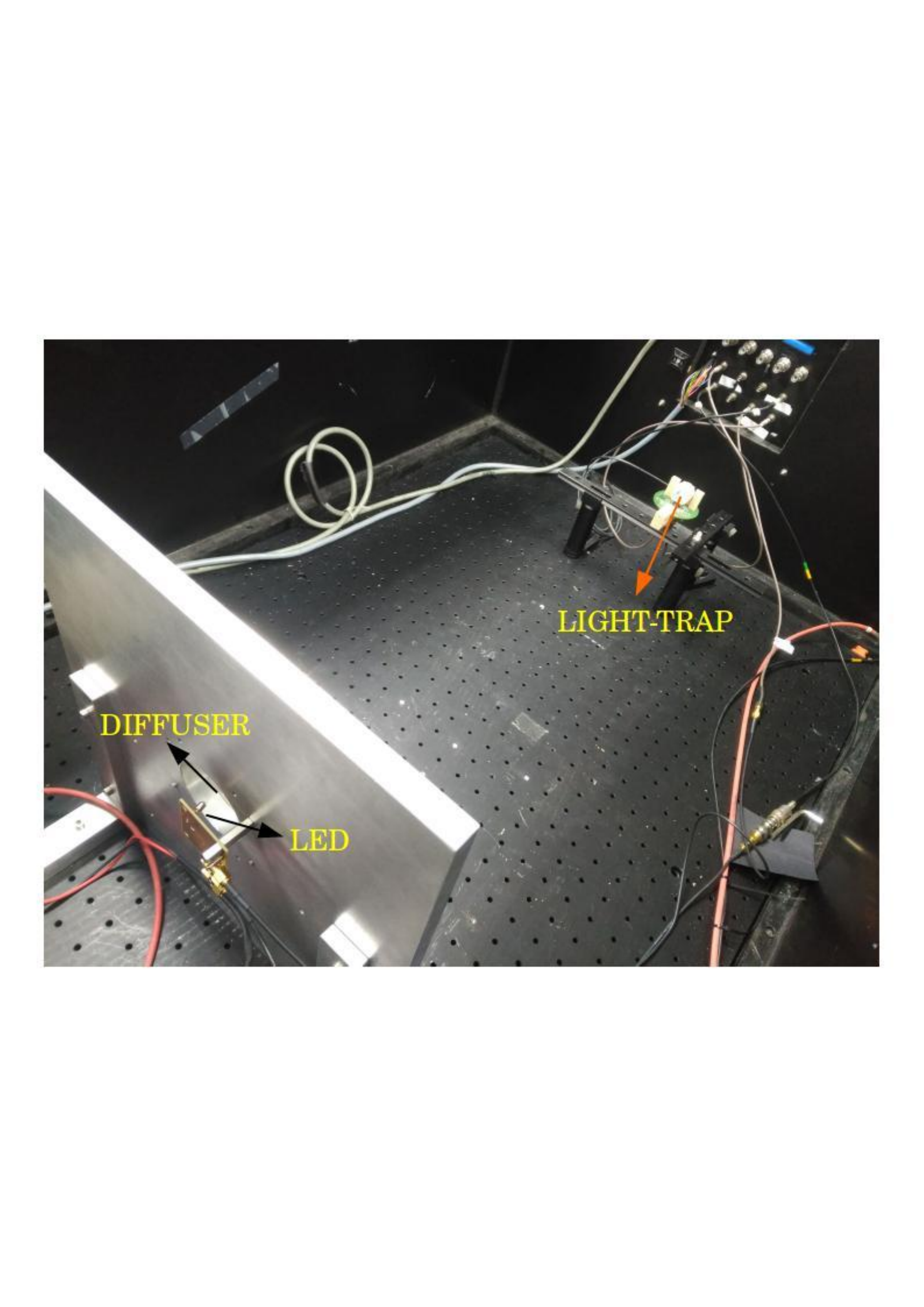}
    \includegraphics[clip=true,trim=0cm 8cm 0cm 8cm,width=0.48\textwidth]{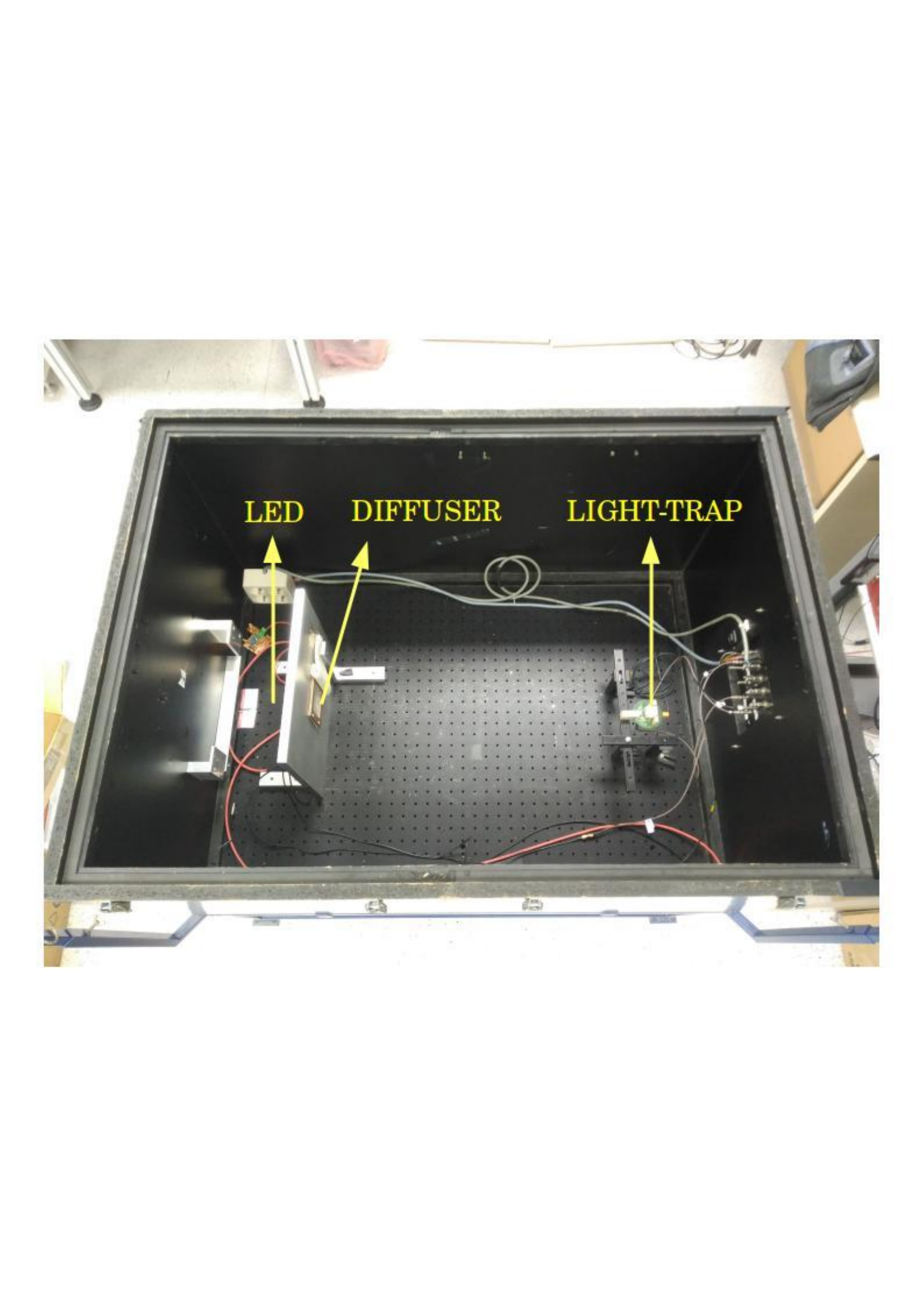}
    \caption{Setup used to characterize the light-trap at the laboratory. All the elements are contained in a dark box. Pulses fired by the LED driver go through the diffuser towards the other side of the box in which the Light-Trap (or the naked SiPM) is set.} 
    \label{fig:LTsetup}
\end{figure}

We tested the system in a dark box (Figure~\ref{fig:LTsetup}) by flashing the Light-Trap with four fast-response LEDs of different colours: $\sim$375, $\sim$445, $\sim$503 and $\sim$600 nm ($\sim$12, $\sim$30, $\sim$30 and $\sim$35~nm FWHM, respectively). Fast pulses of a few ns at a 1 kHz rate are produced by means of a Kaputschinsky LED driver~\citep{Kaputschinsky_2004}. The output signal was recorded using a digital oscilloscope (Rhode\&Schwartz RTO 1024) of 2~GHz bandwidth and 10 GSa/s. Data were stored in 100~ns waveforms from which we can extract the charge, arrival time and other parameters of the detected pulses.

\begin{figure}
    \centering
    \includegraphics[width=0.4\textwidth]{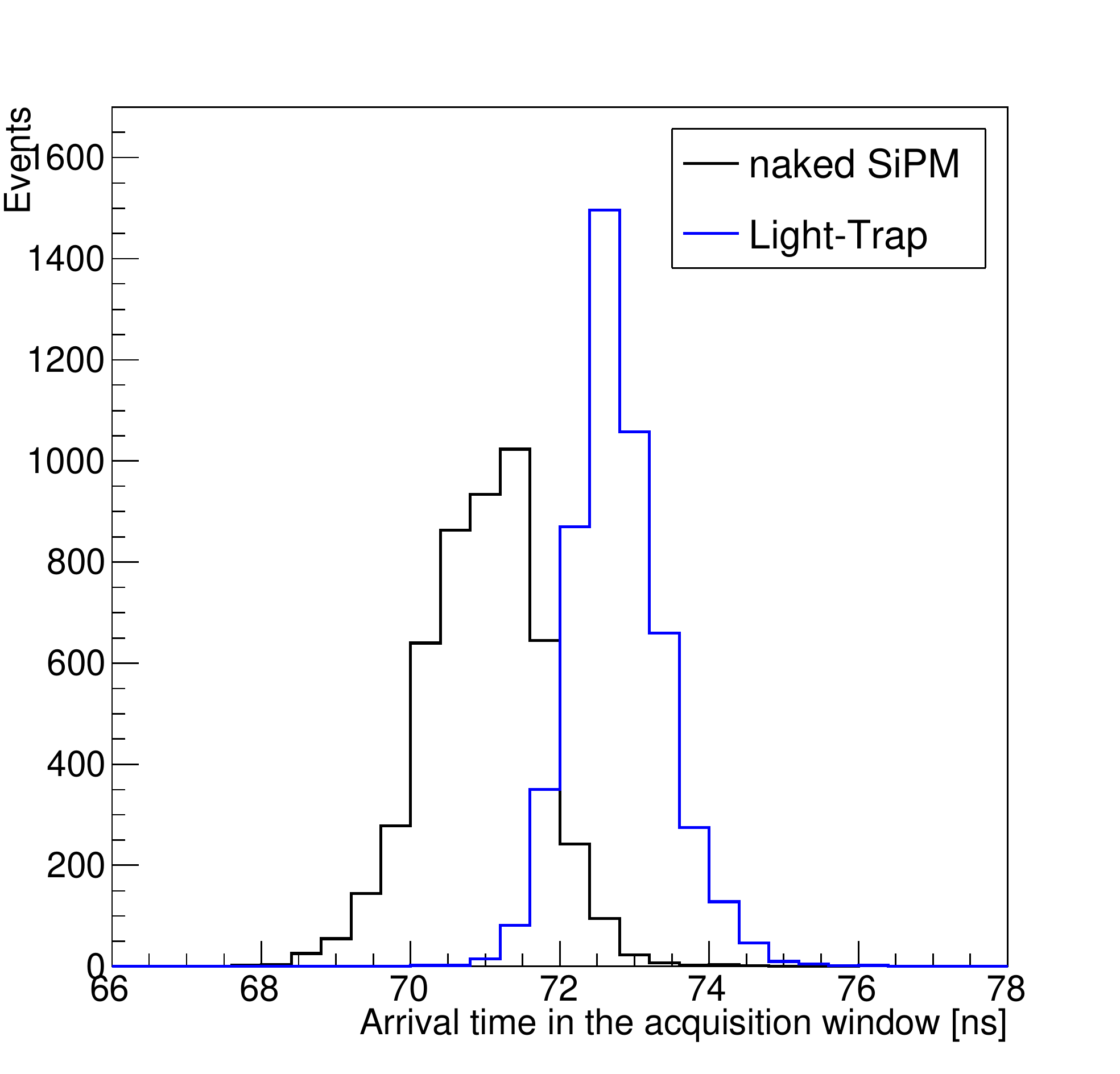}
    \caption{Arrival time distributions of UV pulses for the Light-Trap and for the naked SiPM.}
    \label{fig:LTtiming}
\end{figure}

\begin{figure}
    \centering
    \includegraphics[width=0.8\textwidth]{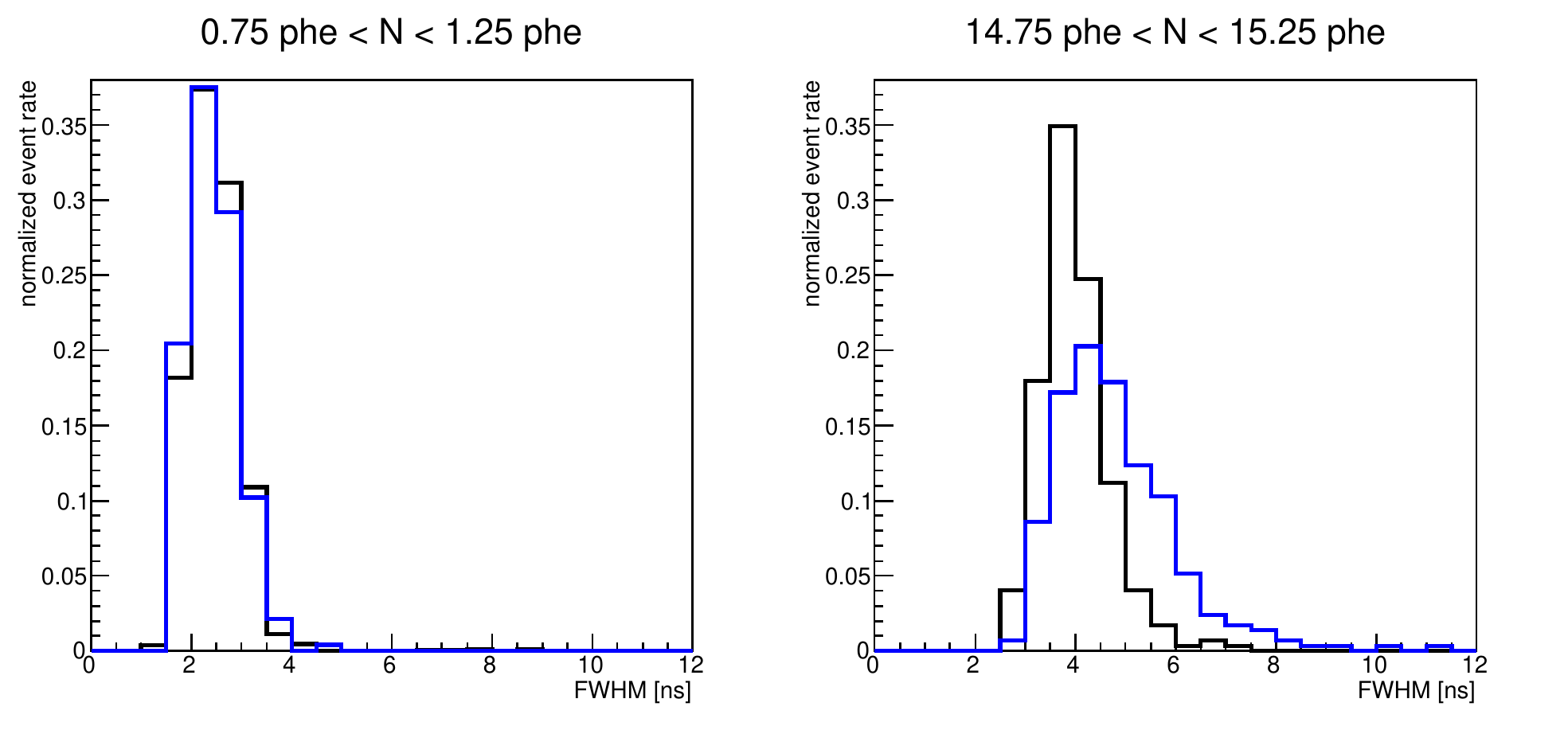}
    \caption{Comparison of the FWHM of UV pulses of 1 (\textbf{left}) and 15 phe (\textbf{right}) events observed with the Light-Trap and with the naked SiPM.}
    \label{fig:LTFWHM}
\end{figure}

The timing properties of the Light-Trap are affected by the re-emission time of the photons absorbed by the WLS and for the total distance that the wavelength-shifted photons travel within the disk before reaching the SiPM. This is observed as a delay in the arrival time and a broadening of the pulses. The delay is observed in Figure~\ref{fig:LTtiming}, which compares the arrival time of 5000 UV pulses detected by the Light-Trap and the naked SiPM. The arrival time here is defined as the time in the recorded window at which the maximum amplitude was found. The delay observed in the Light-trap distribution is of the order of the re-emission time of the WLS. The broadening of the pulses collected with the Light-Trap can be observed in Figure~\ref{fig:LTFWHM}. Pulses of 15~photoelectrons~(phe) recorded with the Light-Trap have a typical FWHM of $\sim$5~ns, while those recorded with the KETEK SiPM are of $\sim$4~ns.

\begin{figure}
    \centering
    \includegraphics[width=0.4\textwidth]{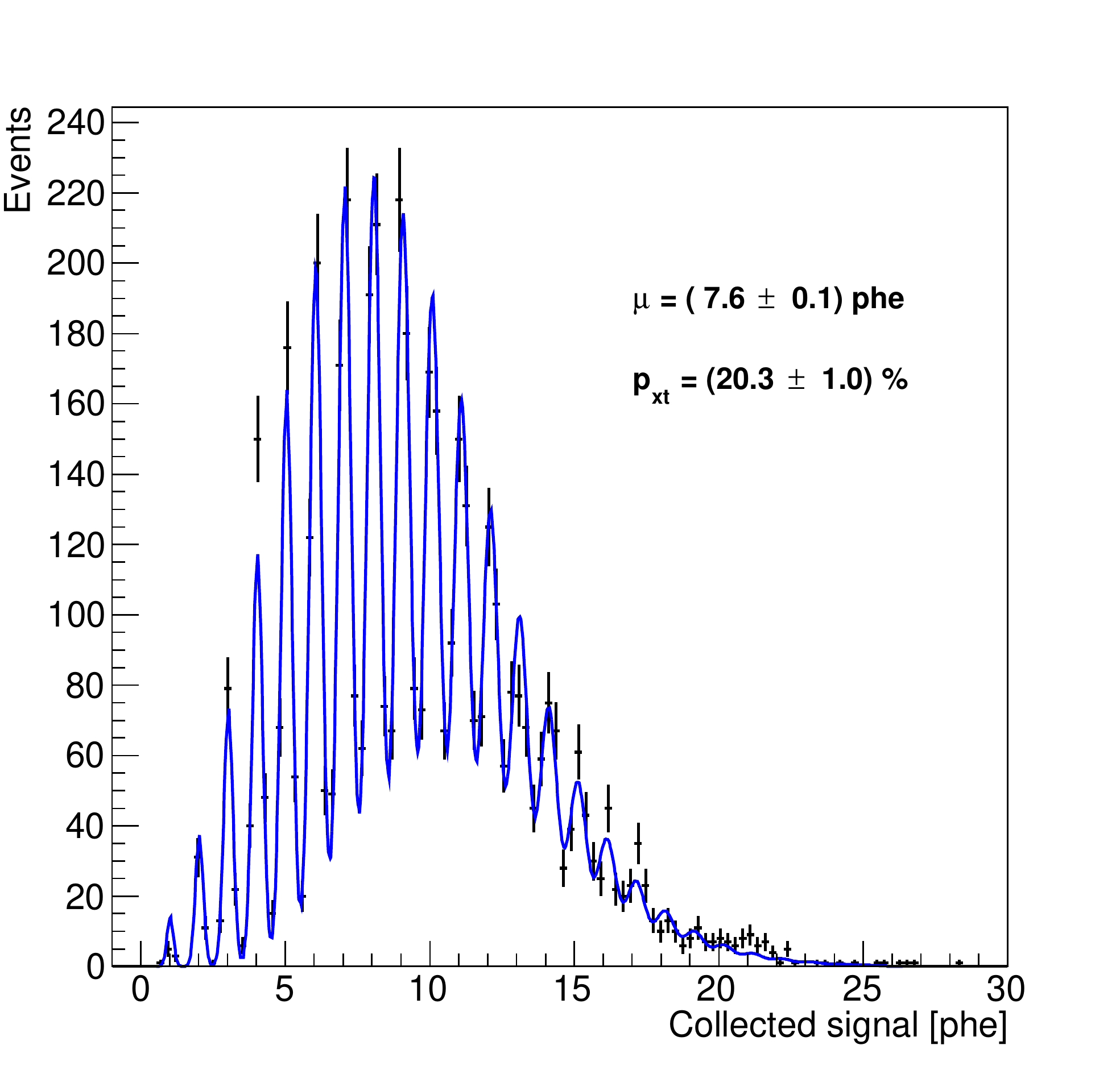}
    \includegraphics[width=0.4\textwidth]{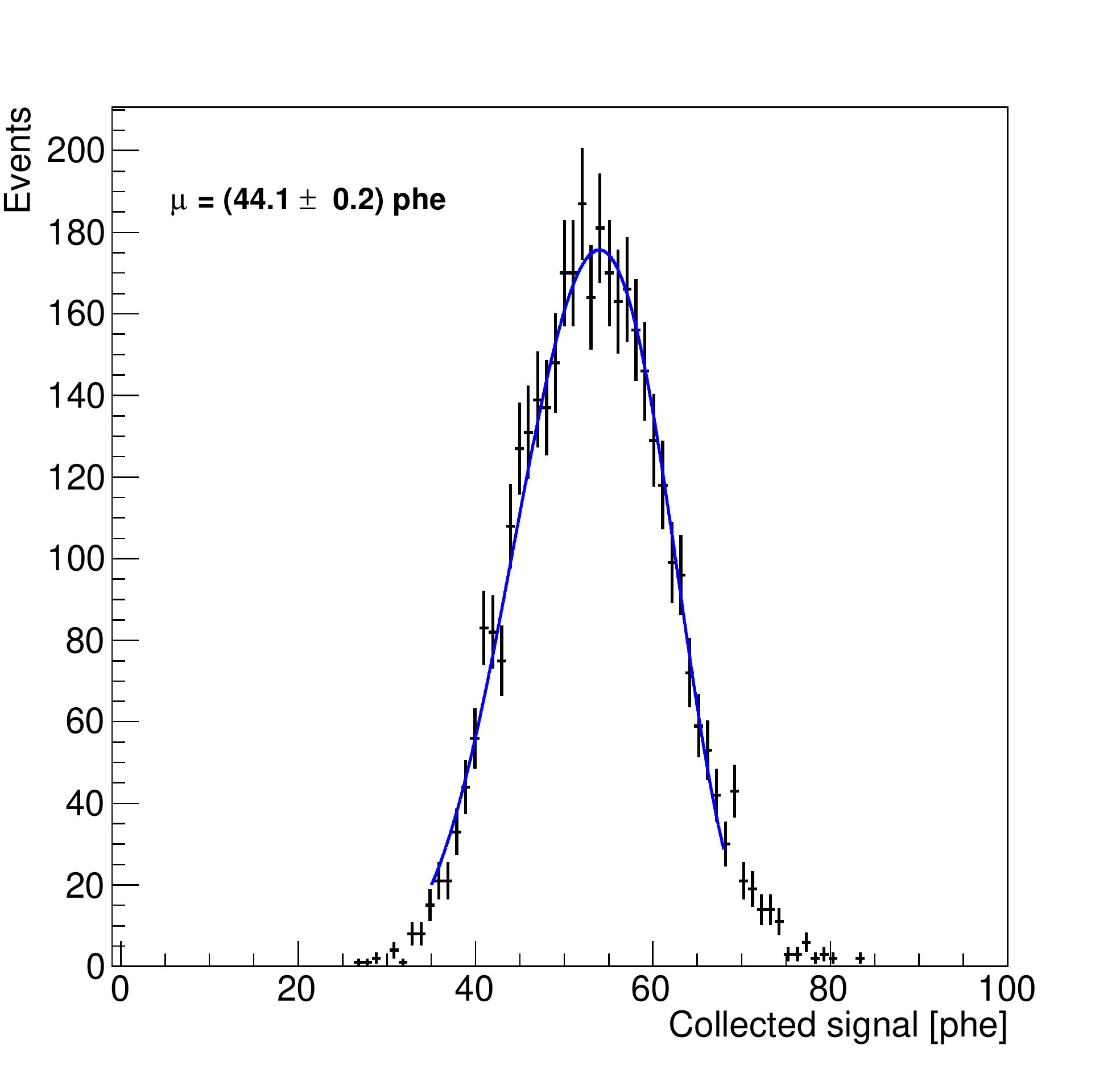}
    \caption{Distributions from data runs with UV illumination ($\sim375$~nm) for naked SiPM (\textbf{left}), and Light-Trap \textbf{(right)}. Blue curve is the fit performed using Eq.~\ref{eq:XTfit}}
    \label{fig:LTmeanPhe}
\end{figure}

To evaluate the efficiency of the Light-Trap on guiding the light towards the SiPM, we compared the mean number of photons $\mu$ detected by the Light-Trap with the mean number of photons detected by the naked SiPM, given the same incident photon flux (Figure~\ref{fig:LTmeanPhe}). When estimating $\mu$ the cross-talk probability $p_{XT}$ must be taken into account. To do so we followed the approach proposed in~\citep{Gallego_2013}, that describes $p_{XT}$ with a binomial distribution:
%The formula can be understood as follows. k successes occur with probability pk and n − k failures occur with probability (1 − p)n − k. However, the k successes can occur anywhere among the n trials, and there are ( n k ) different ways of distributing k successes in a sequence of n trials.

\begin{equation}
    p_{n,m}(p_{XT}) = \left( 1 - p_{XT} \right)^n p_{XT}^{m-n} \binom{m-1}{n-1}
\end{equation}
Then, the probability of detecting a mean number of photons $\mu$ can be written as
\begin{equation}\label{eq:XTfit}
    f(x) = A \sum^N _{n=1} \sum^n _{m=1} p_{n,m}(p_{XT}) P(n\mid \mu) \frac{1}{\sqrt{2\pi}\sigma_n} e^{-\left( \frac{x-n\cdot G + B}{\sqrt{2\pi}\sigma_n} \right)^2} + \text{Ped}(x)
\end{equation}
where $A$ is the normalization, $P(n\mid \mu)$ is the Poisson probability of having $n$ cells fired given a mean number of interacting photons $\mu$, $G$ is the conversion factor from integrated charge to phe, $B$ is the bias or the position of the pedestal peak and $\sigma^2 _n=\sigma^2 _e + n \cdot \sigma^2 _l$ is composed by the electronic noise $\sigma_e$ and the noise related to gain fluctuations $\sigma_l$. $\text{Ped}(x)$ describes the pedestal events as
\begin{equation}\label{eq:Pedfit}
    \text{Ped}(x) = A_0  P(0\mid \mu) \frac{1}{\sqrt{2\pi}\sigma_e} e^{-\left( \frac{x-B}{\sqrt{2\pi}\sigma_{e}} \right)^2}
\end{equation}

with $A_0$ the normalization for this term in the equation. For a detailed explanation on the formalism of treating SiPM data including cross-talk effects refer to~\citep{Gallego_2013} and~\citep{Chmill_2017}. This model does not include explicitly afterpulsing or delayed cross-talk, which could bias the measured flux towards higher values. Although it is hard to distinguish between both type of events, typically delayed cross-talk tends to occur faster than afterpulsing (see figures 19 and 26 in~\citealp{Otte_2017}). Afterpulses mainly happen outside the integration window from which the charge is extracted in our measurements and they should not significantly impact the extracted signal. An afterpulse following a dark count prior to the main light-pulse event (or just a dark count by itself that happened shortly before the arrival of the light pulse) would also affect the charge extraction process. Most of these events are however removed when estimating the baseline, which is done by applying a constant fit to the waveform over the 25~ns preceding the pulse. By applying cuts in the baseline level and in the $\chi ^2$ of such fit it is possible to reject those events in which the SiPM signal is suffering from pile-up from a previous pulse. Many of the delayed cross-talk events may occur inside the integration window from which we extract the charge. In our measurements they are simply not distinguished from fast cross-talk events: the cross-talk probabilities we measured include also a small contribution from delayed cross-talk. The residual contribution from these unwanted events should have a negligible impact considering the precision of a few percent with which we later estimate the efficiency of the Light-Trap.

Equation \ref{eq:XTfit} has eight free parameters, which makes the fit convergence sensitive to the initialization of those parameters\footnote{A simpler method to obtain $\mu$ involving much less parameters is described in~\citep{Otte_2006}. This method is however suitable for low fluxes, with mean values of $\mu$ of a few phe. Our experiments comprise fluxes going from $\sim$1 to $\sim$50 phe.}. To reduce the number of free parameters we first calculate $G$ and $B$ using \textit{dark runs}, measurements performed with the LED switched off and with the trigger level set at $\sim$0.5~phe. A dark run then consists mainly on dark count and dark count + cross-talk events. A dark run is always taken before and after a \textit{data run} (a run under LED illumination) to constantly monitor the SiPM gain. Both dark and data runs are composed of 5000 events. Once the conversion factor is known, the histograms storing data run events are directly built in units of photoelectrons.

The result of fitting $f(x)$ to the spectra obtained with the naked SiPM and with the Light-Trap when flashed with UV light can be seen in Figure~\ref{fig:LTmeanPhe}. In the case of the naked SiPM, where the detected flux is lower, the different multi-electron peaks can be well fitted and distinguished using this model. The value obtained for $p_{XT}$ is consistent with that expected from dark runs, where the cross-talk probability could be roughly estimated as $p_{XT} \simeq \frac{N_{>1.5~\text{phe}}}{N_{>0.5~\text{phe}}} \simeq 20\%$. The obtained parameters for $p_{XT}$, $B$ and $G$ in the naked SiPM histogram are set as fixed parameters for fitting the distributions obtained with the Light-Trap. In those distributions the measured flux is higher and individual peaks are harder to distinguish (and then there is a degeneracy in the parameter space).

\begin{figure}
   \centering
   \includegraphics[width=0.6\columnwidth]{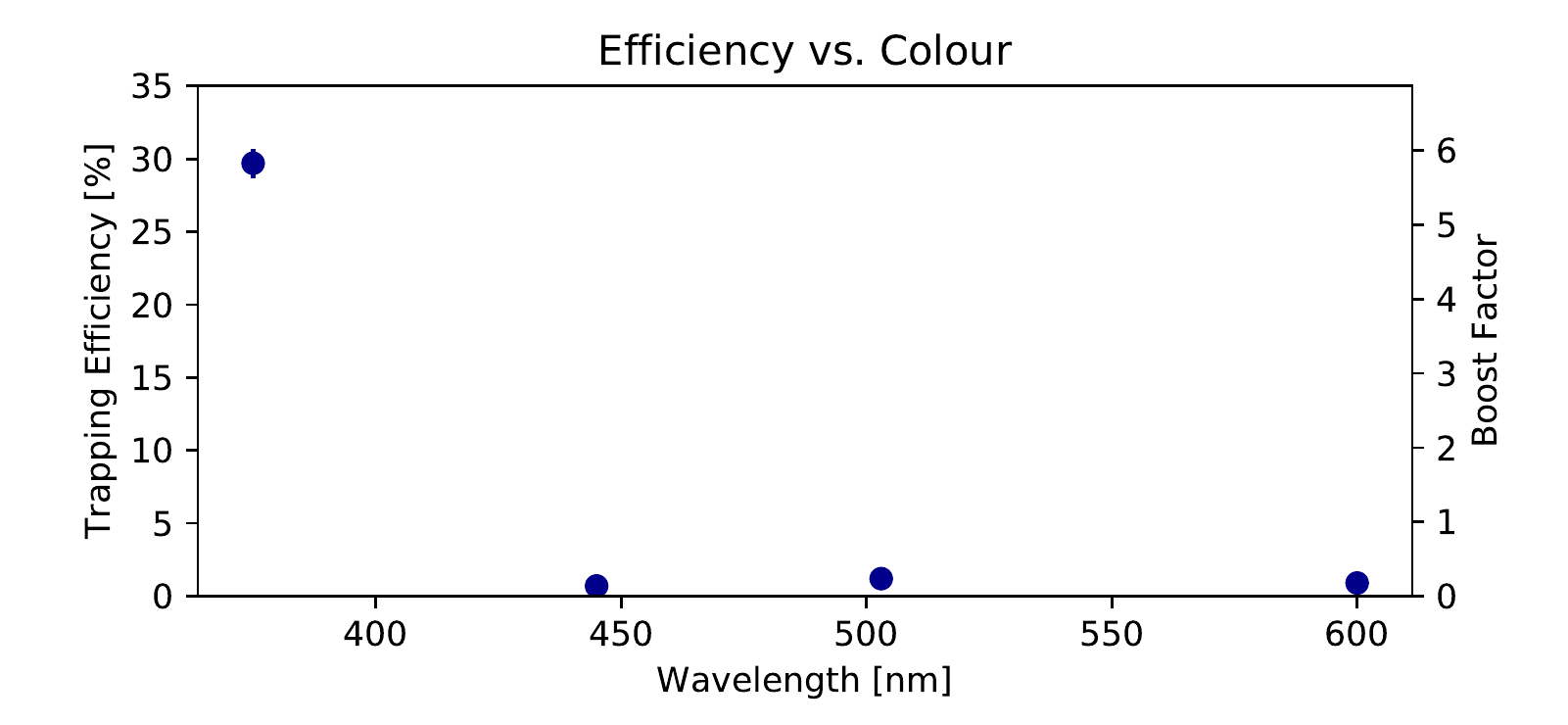} % requires the graphicx package
   \caption{Trapping efficiency and boost factor of the Light-Trap proof of concept pixel as a function of wavelength.}
   \label{fig:EffBoost}
\end{figure}

A simple ratio of the Light-Trap output signal to the one obtained with the naked-SiPM allowed for an estimation of the ``boost factor'' achieved by the additional use of the PMMA disk. The ratio of this boost factor with that expected by the simple geometric consideration (i.e. the increase in area between the 9 mm$^{2}$ SiPM and 176.71 mm$^{2}$ disk, a factor 19.63) gives the ``trapping efficiency'' of the Light-Trap device: the fraction of photons incident in the disk that hit the SiPM. Figure \ref{fig:EffBoost} shows the boost factor and the trapping efficiency of the Light-Trap as a function of wavelength. The PoC pixel collects the same amount of UV light that what would be collected by six SiPMs similar to the one used to build the Light Trap. This also means that the efficiency of the light trap to bring the incident light into the SiPM is $\sim$30\%. The pixel is almost blind to longer wavelengths. 

\section{Monte Carlo simulations}\label{sec:MC}

Several parameters can affect the trapping efficiency of the pixel, such as the distance between the reflective surfaces and the disk, the optical properties of the reflectors, the quality of the optical coupling or the roughness of the disk surface. Most of them are hard to control and manipulate in the PoC pixel described in Section~\ref{sec:pixel_poc}. To understand how these (and other) parameters affect the performance of the Light-Trap we simulated the pixel using the Geant4 software simulation package (version 4.10.01-p02, \citealt{Allison_2006}). The simulated detector consists on a perfectly optically polished 15~mm diameter and 3~mm thick disk with the absorption and emission properties of the EJ-299-15 material. A coupling material (PMMA, with no dopant) of thickness $\Delta$ was also added to optically couple a sensitive detector to the disk (i.e the SiPM, but the properties of the KETEK device have not been included in these simulations). Two mirrors with reflectivity $R$ were placed at distances $d1$ and $d2$ from the sides and bottom of the disk, respectively. We simulated a monochromatic beam of 375~nm photons that arrive perpendicular to the disk surface. Each run consists of 5000 events.

In a first set of simulations we aimed at understanding the impact that the different Light-Trap elements have in the overall trapping efficiency. Table \ref{tab:overall} shows the impact that $d_1$, $d_2$, $\Delta$ and $R$ have on the trapping efficiency $\epsilon$. Entry $A.1$ represents the close-to-ideal scenario in which mirrors are placed at a negligible distance from the disk walls and $\Delta \rightarrow 0$ (note that since the disk is a cylinder and the detector is flat $\Delta=0$ is fulfilled only at the centre of the coupling and it increases toward the edges, from which some photons could still escape). Assuming a quantum yield of $Y=100\%$ the efficiency obtained can reach $\sim$69\%, which is consistent with expectations from geometrical calculations plus Fresnel losses. This is the highest efficiency that could be obtained if we could control and optimize all the parameters. As we deviate from this ideal scenario, the efficiency drops. Parameters like $d_1$ and $R$ have a strong impact ($A.2$ and $A.4$ in Table~\ref{tab:overall}, while $d_2$ does not seem to significantly affect $\epsilon$ ($A.3$). The effect of $\Delta$ seems to be not so critical ($A.5$) but it should actually be treated with caution. In the simulation the coupling is a prism with perfectly polished walls in which up to $\sim$30\% of the detected photons first experienced TIR ($\Delta=0.8$~mm). This may not represent accurately what happens for instance with the silicone rubber sheet described in Section~\ref{sec:coupling}, where it is hard to predict what happens to the photons reaching the interface silicone-air. Thus, it is important to keep the thickness of the optical coupling as low as possible. Finally, entry $A.6$ of Table~\ref{tab:overall} evidences the importance of adding the reflective surfaces to recover a fraction of the photons that escape through the bottom and side-walls.

\begin{table}
    \centering
    \begin{tabular}{ c || c c c c || c | c }
         Nr & $d_1$ & $d_2$ & R &  $\Delta$ & $\epsilon$ (Y=100\%) & $\epsilon$ (Y=84\%) \\
          & [$\mu$m] & [$\mu$m] & [\%] & [mm] & [\%] & [\%] \\
         \hline \hline
         A.1 & 1 & 1 & 100 & 0 & 69 & 58 \\
         A.2 & \textbf{100} & 1 & 100 & 0 & 56 & 47 \\
         A.3 & 1 & \textbf{100} & 100 & 0 & 69 & 58 \\
         A.4 & 1 & 1 & \textbf{98} & 0 & 59 & 50 \\
         A.5 & 1 & 1 & 100 & \textbf{0.8} & 65 & 55 \\
             \hline
         A.6 & - & - & \textbf{0} & 0 & 28 & 23 \\
         \hline
    \end{tabular}
    \caption{Simulations performed with Geant4. $d_1$ and $d_2$ are the distance between disk and side and bottom mirrors, respectively, $R$ is the reflectivity of the mirrors and $\Delta$ is the thickness of the coupling. The trapping efficiency $\epsilon$ is computed por quantum yields $Y$ of 100 and 84\%.}
    \label{tab:overall}
\end{table}

In a second set of simulations we attempted to understand the measurements in Section~\ref{sec:LT_poc} using parameters closer to the PoC pixel. Table~\ref{tab:sys} shows the results. For a quantum yield of 84\%, $\epsilon$ ranges  between 33 and 44 \%, above the efficiency measured at the laboratory (Section~\ref{sec:lab}). There are a few aspects in which simulations certainly do not describe the PoC pixel. One is the previously mentioned effect of photons experiencing TIR in the coupling material. A second one has to do with the way in which the mirrors and the disk are placed in the holder. In the simulations the PMMA disk is floating in air. In the PoC pixel the two plastic screws of the pixel holder that push the disk towards the SiPM (Section~\ref{sec:holder}) deform the reflective foil on the sides, so that it touches the disk near the screws and separates itself in other regions. The effective reflectivity is then not homogeneous all around the disk. Besides, we are assuming that the reflective surfaces, the disk and the coupling material have no roughness.

Surface roughness can prevent TIR or randomize the angle of photons when they cross the surface, both increasing losses. Even if we have crosschecked that roughness is not significant for photons crossing twice the flat surface of the disk (Section~\ref{sec:disk}), a small roughness under the sensitivity of our measurement can multiply its effect after several interactions with the disk walls. Besides, we have not studied the curved sides of the disk. A full Monte Carlo evaluation of the PoC light trap goes beyond the goal of this paper. We consider that these Monte Carlo results are in reasonable agreement with the measurements.

\begin{table}[htb]
    \centering
    \begin{tabular}{ l || l | l | l || c | c }
         Nr & d1 & R &  $\Delta$ & $\epsilon$ (Y=100\%) & $\epsilon$ (Y=84\%) \\
          & [$\mu$m] & [\%] & [mm] & [\%] & [\%] \\
         \hline \hline
         B.1 & \multirow{6}{1cm}{100} & \multirow{2}{1cm}{100} & 0.5 & 52  & 44  \\
         B.2 &  & & 0.8 & 49 & 41 \\ \cline{3-6}
         B.3 &  & \multirow{2}{1cm}{98} & 0.5 & 48 & 41 \\
         B.4 &  & & 0.8 & 47 & 39 \\ \cline{3-6}
         B.5 &  & \multirow{2}{1cm}{96} & 0.5 & 45 & 38 \\
         B.6 &  & & 0.8 & 44 & 37 \\
        \hline
        B.7 & \multirow{6}{1cm}{300} & \multirow{2}{1cm}{100} & 0.5 & 45 & 38 \\
        B.8 &  & & 0.8 & 44 & 37 \\ \cline{3-6}
        B.9 &  & \multirow{2}{1cm}{98} & 0.5 & 43 & 36 \\
        B.10 &  & & 0.8 & 41 & 35 \\ \cline{3-6}
        B.11 &  & \multirow{2}{1cm}{96} & 0.5 & 41 & 34 \\
        B.12 &  & & 0.8 & 39 & 33 \\
        \hline
    \end{tabular}
    \caption{Simulations performed with Geant4. All the entries were computed with $d_2 = 100$~$\mu$m.}
    \label{tab:sys}
\end{table}

\section{Conclusions}

We presented a new concept of photodetector that aims to provide a non-expensive solution to cover a large detection area, while profiting from the advantages SiPMs provide (single-photon resolution, low voltage operation, robustness...). This solution relies on the use of a PMMA disk doped with a wavelength shifter material to ``trap'' light over the whole are of the disk and guide it to the SiPM. With an standardized disk production, the cost of the pixel should be dominated by the price of the SiPM.

We built a proof-of-concept pixel and tested it on the lab. The measurements have shown that the Light-Trap concept works: light is collected only in the desired wavelength range over a detection area that is much larger than what can be achieved with a single SiPM. The trapping efficiency is modest, but there are several actions that could be performed to improve its performance. From the simulations performed with Geant4 it is clear that the reflectivity of the mirrors and especially their distance to the disk are critical (and minimizing that distance is particularly challenging). Reducing the thickness of the coupling material would also help to improve the efficiency. By using other adhesives that directly glue the SiPM to the disk we would lose the possibility of removing the light sensor for testing, but could deposit thinner layers. Also using a square-like PMMA block instead of a disk would facilitate the coupling.

With an improved trapping efficiency, the concept of the Light-Trap offers a suitable alternative to build a low-cost large-area detector for IACT astronomy, and for astroparticles physics and high energy physics in general. Especially for those applications where the efficiency loss can be compensated with an increase in the detection area. A good example are new generation neutrino detectors, which feature large volumes. In fact, similar light-trapping schemes have been proposed for IceCube~\citep{Hebecker_2016} or DUNE (ARAPUCA, ~\citealp{Machado_2016}). Other methods employ complex optical designs to increase the detection area (see for instance \citealt{Vassiliev_2007}). With this solution a higher sensitivity can be obtained, but the complexity is normally higher, it is more difficult to scale to larger sizes and the efficiency of such systems depend strongly on the incident angle of the light (while the Light-Trap allows to collect light practically from any angle).

The applications of the Light-Trap may not be limited only to science, but could also bring new opportunities in industry. With proper modifications it could be suitable for instance for Single Photon Emitted Computer Tomography (SPECT) systems and particularly in the veterinary field, where a cost reduction is often more demanded than an improvement in sensitivity. With a relatively large pixel and a simple and compact readout electronics the amount of shielding material needed to build a SPECT camera could be significantly reduced, and hence also its weight and price~\citep{Peterson_2011}. By choosing the adequate dopant, the Light-trap can be tuned to achieve a sensitivity in a desired wavelength range, suitable for its specific application.

\section*{Acknowledgments}
This work would not have been possible without the support of the engineering and technical staff at IFAE, especially  Joan Boix, Javier Gaweda and Jos\'e Illa. The authors would also like to acknowledge the fruitful discussions undertaken with Eljen representatives. The project was supported by a Marie Sklodowska-Curie individual European fellowship (EU project 660138 -- Light-Trap), Centro de Excelencia Severo Ochoa grant SEV-2012-0234 and by the Otto Hahn Award of the Max Planck Society (PI: D. Mazin).

\section*{References}

\bibliography{mybibfile}

\end{document}